\documentclass[a4paper]{article}

\usepackage[fussy]{INTERSPEECH2021}
\usepackage{tipa} %
\newcommand{\ipa}{\textipa}
\usepackage[
        style=ieee,
        doi=false,
        eprint=false,
        url=false,
        maxbibnames=3
    ]{biblatex}
\bibliography{refs.bib}
 \defbibheading{header}[\refname{\section{References}}]

\title{Speech Representations and Phoneme Classification\protect\\ for Preserving the Endangered Language of Ladin}

\name{Zane Durante$^{*,1}$\thanks{$^{*}$equal contribution, alphabetical order}, Leena Mathur$^{*,1,2}$, Eric Ye$^{1,2}$, Sichong Zhao$^{1,2}$, Tejas Ramdas$^1$, Khalil Iskarous$^{2}$}
\address{
  $^1$Center for Artificial Intelligence in Society 
  \hspace{5pt}
  $^2$Department of Linguistics\\
  University of Southern California, Los Angeles, USA}
\email{\{durante, lmathur, eye, sichongz, tramdas, kiskarou\} @ usc.edu}

\begin{document}

\maketitle
\begin{abstract}
  A vast majority of the world's 7,000 spoken languages are predicted to become extinct within this century, including the endangered language of \textit{Ladin} from the Italian Alps. Linguists who work to preserve a language's phonetic and phonological structure can spend hours transcribing each minute of speech from native speakers. To address this problem in the context of Ladin, our paper presents the first analysis of speech representations and machine learning models for classifying 32 phonemes of Ladin. We experimented with a novel dataset of the Fascian  dialect of Ladin, collected from native speakers in Italy. We created \textit{frame-level} and \textit{segment-level} speech feature extraction approaches and conducted extensive experiments with 8 different classifiers trained on 9 different speech representations. Our speech representations ranged from traditional features (MFCC, LPC) to features learned with deep neural network models (autoencoders, LSTM autoencoders, and WaveNet). Our highest performing classifier, trained on MFCC representations of speech signals, achieved an 86\% average accuracy across all Ladin phonemes.  We also obtained average accuracies above 77\% for all Ladin phoneme subgroups examined. Our findings contribute insights for learning discriminative Ladin phoneme representations and demonstrate the potential for leveraging machine learning and speech signal processing to preserve Ladin and other endangered languages. 
  %
\end{abstract}
\noindent\textbf{Index Terms}: speech representations,  phoneme classifiers, machine learning, language preservation, speech signal processing

\section{Introduction}
\label{sec:intro}
Linguists estimate that 60\% to 90\% of the world’s 7,000 spoken languages are likely to become extinct within the next century \cite{doi:10.1111/j.1749-818X.2007.00004.x}.
The extinction of any language represents an \textit{irreversible} loss of information for humanity, as unique linguistic, psychological, and sociocultural information are embedded within the phonology, syntax, and semantics of each language \cite{10.2307/416368}. The pressing nature of this problem has motivated linguists to conduct research and community interventions to preserve endangered languages \cite{doi:10.1111/j.1749-818X.2007.00004.x, grenoble_whaley_2005}. Linguists begin language preservation efforts by collecting spoken data from native speakers and transcribing \textit{phonemes}, the fundamental speech sounds and linguistic units of a language \cite{10.3366/j.ctt1g09vvm, 2004}. Since each minute of speech can take up to 2 hours for a trained linguist to transcribe \cite{adams18evaluating}, and each hour can take around 100 hours to transcribe \cite{cavar-etal-2016-endangered}, manual phoneme classification represents a significant bottleneck in endangered language preservation, motivating the development of computational approaches to assist humans in this task. 

Advances in machine learning and speech signal processing have demonstrated the effectiveness of deep neural network models for learning discriminative speech representations \cite{latif2020deep} and classifying phonemes \cite{1556215}, when trained on large speech datasets. However, existing methods for automated phoneme classification have been largely ineffective for endangered languages \cite{adams18evaluating}, due to the limited number of speakers and the difficulty in collecting labeled, transcribed training data. 
Our paper contributes to endangered language preservation efforts in the specific context of \textit{Ladin}, an endangered language from the Italian Alps. We use a novel dataset of Ladin collected during linguistic field research in Italy with native Ladin speakers. The Ladin people live in multiple valleys around the \textit{Sella} massif mountain range, and the speakers across valleys exhibit a considerable amount of linguistic variation: this research focuses on classifying phonemes in the  \textit{Fascian} dialect of Ladin, spoken in the Fassa Valley \cite{yang_walker_vietti_chiocchetti_2021}. 

To the best of our knowledge, this paper presents the first attempt to develop computational models that classify Ladin phonemes: we contribute a novel analysis of speech representations and machine learning models for classifying 32 phonemes of Ladin \cite{yang_walker_vietti_chiocchetti_2021}. Given the extremely small amount of labeled phoneme data available for endangered languages, including Ladin, our research was driven by the following two questions: (1) Will classical machine learning models outperform deep neural networks when classifying the phonemes of Ladin? (2) Should phoneme classifiers for Ladin be trained on traditional speech representations (e.g., MFCC, LPC)  or representations learned by deep unsupervised models, such as autoencoders? 

To address these questions, we experimented with 7 classical machine learning models and 1 dense neural network trained on 9 different speech representations that were extracted with traditional algorithms (MFCC, LPC) and autoencoder-based approaches (autoencoder, LSTM autoencoder, WaveNet). Our highest performing model (dense neural network with MFCC representations) achieved an 86\% accuracy across all phonemes, accuracies above 90\% on 6 individual phonemes, and accuracies above 77\% across Ladin phoneme subgroups.



For the endangered language of Ladin, this paper makes two major contributions towards computational approaches that assist linguists in endangered language preservation: (1) We present the first automated phoneme classification approach for Ladin. (2) We present an extensive analysis of speech feature extraction algorithms, speech representations, and phoneme classification models for Ladin. Our overall multi-stage research initiative to preserve Ladin is conceptualized in \textbf{Figure 1}. 

\begin{figure}[h]
    \includegraphics[width=\linewidth]{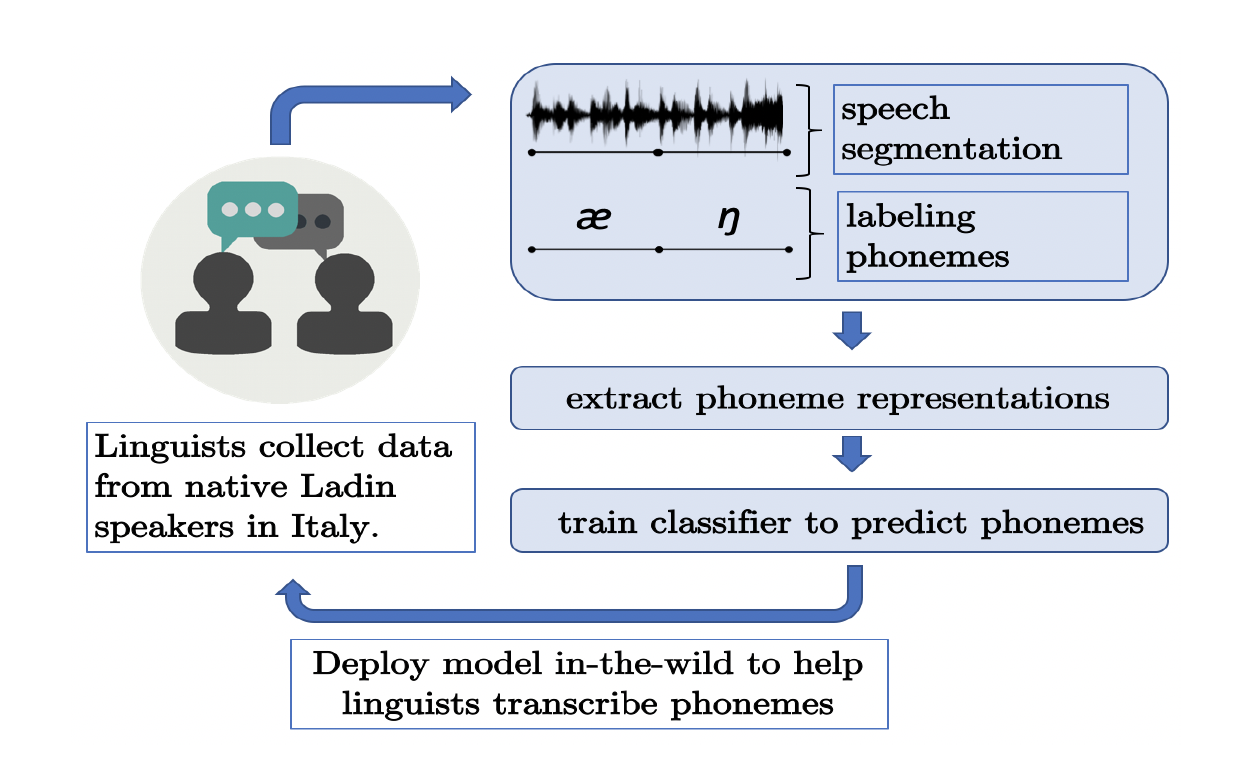}
    \label{fig:my_label}
     \vspace{-0.02\textheight}
    \caption{Multi-stage research process to preserve the Ladin language. This paper focuses on the phoneme representation and classification phases.}
\end{figure}

\section{Background}
\subsection{Endangered language speech research}
\label{ssec:computational methods}
Phoneme classification has largely relied on classical machine learning models (e.g., Support Vector Machines) trained on acoustic waveform representations \cite{inproceedings}, recurrent neural networks (e.g., LSTM) \cite{279192, 1556215}, and convolutional neural networks (e.g., WaveNet) \cite{10.1109/TASLP.2019.2938863}. However, a majority of prior approaches have assumed the existence of hundreds of hours of training data, which is infeasible to collect for low-resource, endangered languages that sometimes only have a few minutes of labeled data available, if at all \cite{adams18evaluating, tsunoda2006language}. Researchers have developed several approaches to address this challenge \cite{computational, anastasopoulos2019}. Speech recognition systems for low-resource languages have shown improvement through data augmentation \cite{augmentation, ragni2014data} and pre-training on high-resource languages \cite{pretraining}. Multilingual models sharing cross-lingual parameters have shown potential in classifying phonemes of the endangered languages of Inuktitut and Tusom \cite{li20icassp}.  A recent transcription tool for linguists, \textit{Persephone} \cite{article}, leveraged recurrent networks trained on limited amounts of the endangered languages of Na (224 minutes) and Chatino (50 minutes).

\subsection{Speech representations}
\label{ssec:representation learning}
A variety of approaches for extracting discriminative representations of speech waveforms have been developed from decades of speech signal processing research \cite{inbook}.
Mel-frequency cepstral coefficients (MFCC), based on cepstral information and the Mel frequency scale, are one of the most commonly used speech representations \cite{5947590}. 
Linear Predictive Coding (LPC) approximates formants by applying linear prediction on the spectral domain \cite{inbook}. More recently, deep neural networks have demonstrated considerable potential for learning useful speech representations \cite{10.1109/TASLP.2019.2938863, DBLP:conf/interspeech/ChungHTG19}. WaveNet, a deep generative model \cite{oord2016wavenet}, has been leveraged to learn discriminative representations of  music and speech \cite{10.1109/TASLP.2019.2938863, nsynth2017}.

\section{Methodology}
\label{sec:methodology}
Our approach involved the following steps: (1) collecting data from native Ladin speakers, (2) extracting traditional speech representations, (3) extracting autoencoder speech representations, and (4) experimenting with machine learning models that were trained on both types of representations to classify Ladin phonemes. 

\subsection{Ladin dataset collection from native speakers}
\label{ssec:dataset} 
Our Ladin dataset was collected through 3 months of field research with native Ladin speakers in the Fassa Valley of northern Italy. Our data consist of 76 audio files that span a total of 5 hours and 38 minutes of conversations with four native speakers of Ladin, (three female, one male). The audio was recorded at 44,100 Hz. Teams of linguists trained in the Ladin language were recruited to transcribe the phonemes of two interviews (one female speaker and one male speaker) with a combined duration of 7 minutes and 55 seconds. There are a total of 2520 labeled phoneme segments across both labeled interviews, with each label belonging to one of 33 phoneme classes: 10 vowel classes (39\% of the audio), 22 consonant classes (47\% of the audio), and 1 class representing the absence of a phoneme (14\% silent segments). Vowels spanned two categories with the following distributions: rounded (8\%) and unrounded (31\%). Consonants spanned 6 key categories with the following distributions: affricates (2\%), approximants (8\%), fricatives (8\%), nasals (7\%), plosives (17\%), and trills (6\%). 

\subsection{Traditional speech representations}
\label{ssec:representations}
\begin{figure}[h]
    \vspace{0.02\textheight}
    \includegraphics[width=\linewidth]{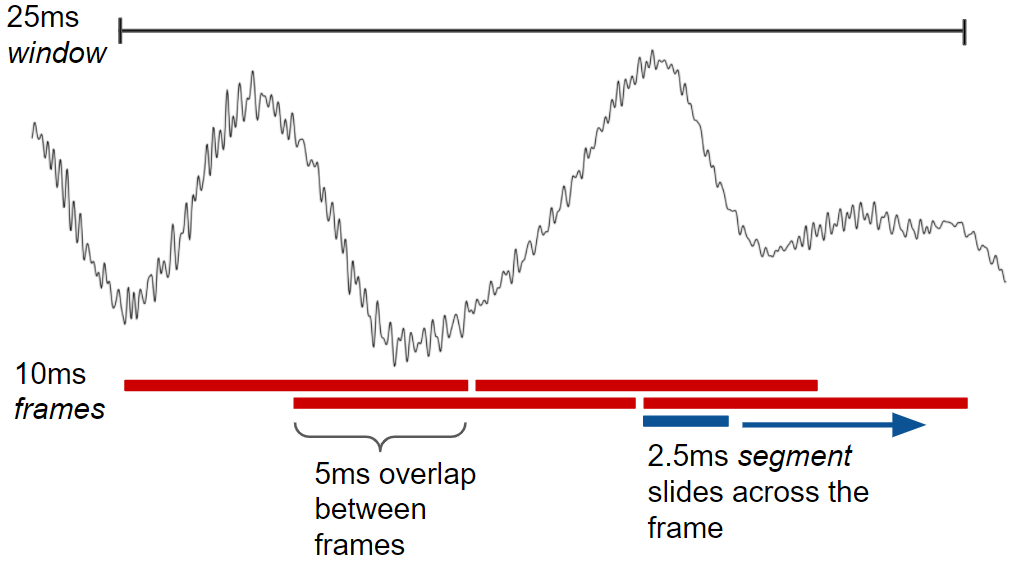}
        \caption{The 25ms audio window is broken up into four 10ms audio frames, each overlapping by 5ms.  For the \textit{frame-level} approach, features are extracted from each frame.  For the \textit{segment-level} approach, each frame is evaluated by sliding a 2.5ms segment across the frame to calculate the average segment-level features per frame. This sample audio is taken from our dataset, and the line segments are drawn to scale.}
    \label{fig:frame_window_segment_diagram}
\end{figure}

To represent raw speech, we extracted speech features with the traditional algorithms of MFCC and LPC \cite{brian_mcfee_2020_3955228}. Given a window size of 25ms that begins at the start of a phoneme, we experimented with two feature extraction approaches, detailed below, for computing speech representations from this window of audio: (1) frame-level extraction and (2) segment-level extraction.  \textbf{Figure \ref{fig:frame_window_segment_diagram}} visualizes the relationship between the audio windows, frames, and segments.

\subsubsection{Frame-level extraction}
\label{sssec:frame extraction}
The \textit{frame-level} extraction approach divided the 25ms audio window into 4 smaller frames of size 10ms, overlapping each other by 5ms. Features for each 10ms frame were then obtained using MFCC and LPC. Next, features obtained from each frame were concatenated to obtain the \textit{frame-level} representation of the 25ms audio window. For each frame, 12 coefficients were extracted for MFCC, and 8 coefficients were extracted for LPC.  


\begin{table*}[ht]
\caption{Average classification accuracy computed across all Ladin phonemes for each of 8 classifiers trained on 9  different traditional and autoencoder speech representations. The best results for each classifier are bolded, and * denotes the highest classification accuracy out of all experiments. }
\centering
\begin{tabular}{l|c|c|c|c|c|c|c|c}
\multicolumn{1}{l|}{\textbf{Representation}} & \multicolumn{1}{l|}{\textbf{Dense NN}} & \multicolumn{1}{l|}{\textbf{SVM (linear)}} & \multicolumn{1}{l|}{\textbf{SVM (rbf)}} & \multicolumn{1}{l|}{\textbf{RF}} & \multicolumn{1}{l|}{\textbf{DT}} & \multicolumn{1}{l|}{\textbf{LR (L1)}} & \multicolumn{1}{l|}{\textbf{LR (L2)}} & \multicolumn{1}{l}{\textbf{LR (ElasticNet)}} \\ \hline
\multicolumn{1}{c}{\textbf{Traditional}} \\ \hline
MFCC Frame & 0.82 & 0.29 & 0.43 & 0.51 & 0.46 & 0.46 & 0.46 & 0.47 \\
MFCC Segment & \textbf{0.86*} & 0.36 & 0.52 & \textbf{0.58} & 0.52 & 0.52 & 0.53 & \textbf{0.56} \\
LPC Frame & 0.67 & 0.23 & 0.36 & 0.37 & 0.35 & 0.35 & 0.35 & 0.35 \\
LPC Segment & 0.31 & 0.15 & 0.18 & 0.19 & 0.18 & 0.18 & 0.18 & 0.19 \\
\hline
\multicolumn{1}{c}{\textbf{Autoencoder}} \\ \hline
Small AE & 0.54 & 0.28 & 0.39 & 0.33 & 0.37 & 0.36 & 0.33 & 0.38 \\
Big AE & 0.50 & 0.25 & 0.34 & 0.31 & 0.34 & 0.34 & 0.32 & 0.36 \\
Small LSTM AE & 0.53 & 0.28 & 0.39 & 0.33 & 0.37 & 0.36 & 0.33 & 0.38 \\
Big LSTM AE & 0.35 & 0.10 & 0.16 & 0.18 & 0.17 & 0.17 & 0.17 & 0.17 \\
WaveNet & 0.77 & \textbf{0.39} & \textbf{0.59} & 0.52 & \textbf{0.56} & \textbf{0.54} & \textbf{0.57} & 0.52
\end{tabular}
\end{table*}

\subsubsection{Segment-level extraction}
\label{sssec:segment extraction}
The \textit{segment-level} extraction approach used the same 10ms audio frames as the \textit{frame-level} approach. However, for each 10ms frame, features were calculated using a smaller 2.5ms (110 bits) segment that slides across the frame a single bit at a time, creating segments that overlap by 109 bits. Features for each 2.5ms segment were computed and averaged across the frame.  Then, the average segment-level features for each frame were concatenated to form the \textit{segment-level} representation.  Similar to \textit{frame-level} extraction, 12 coefficients were extracted for MFCC, and 8 coefficients were extracted for LPC. This approach was informed by prior research using Restricted Boltzmann Machines to represent raw speech \cite{5947700}.

\subsection{Autoencoder speech representations}
\label{ssec:wavenet extraction}
Our experiments used traditional autoencoders, LSTM autoencoders, and a WaveNet autoencoder to represent raw speech.

To establish a baseline for how the unlabeled audio files could be leveraged to learn speech representations, we trained two traditional autoencoders (AEs) on the unlabeled Ladin audio, denoted as \textit{Small AE} and \textit{Big AE} with bottleneck layer sizes of 8 and 16, respectively. To explore the possibility of leveraging temporal information encoded in our data, we also trained two LSTM autoencoders \cite{10.5555/3045118.3045209} on the unlabeled Ladin audio, denoted as \textit{Small LSTM AE} and \textit{Big LSTM AE} with bottleneck layer sizes of 8 and 16, respectively. All models took audio tensor inputs of length 110 (corresponding to 2.5ms of audio data). The initial hidden layer for all models had 2048 neurons, with each subsequent encoding layer having half as many neurons as the previous layer, until the bottleneck layer. The decoding layers mirror each corresponding encoding layer in size. All models used the ReLU activation function on all layers except the final layer, which used tanh, since the audio is bounded on $[-1,1]$. 
The bottleneck representations from all autoencoders were extracted per the \textit{segment-level} approach in 3.2.2.

The \textit{WaveNet} extraction approach used a WaveNet autoencoder \cite{nsynth2017} trained on 306,043 sounds of different pitch, timbre, and envelope, which are common properties of human speech. Our approach was inspired by prior WaveNet autoencoders that learned discriminative  representations of English, French, and Mandarin phonemes  \cite{10.1109/TASLP.2019.2938863}. The WaveNet autoencoder extracts 16-dimensional encodings from each 512-bit phoneme speech segment. Phoneme segments shorter than 512 were padded with zeros to become valid inputs. Phoneme segments longer than 512 bits were divided into 512-bit segments which were fed into the autoencoder. Encodings were then averaged to obtain the final 16-dimensional speech representation. 

\subsection{Classification experiments}
\label{ssec:experiments}

Phoneme transcription was formulated as a multi-class classification problem to assign each audio segment to one of 33 classes (32 labels for phonemes and 1 label for silent speech segments). To identify effective combinations of speech representations and classification algorithms for Ladin phoneme transcription, we experimented with 8 machine learning classifiers trained on 9 different speech representations. We tested a 2-layer dense neural network (Dense NN, 512 neurons per layer, dropout of 0.05, adam optimizer, 50 epochs), implemented with Keras. 
We also implemented the following classifiers with the scikit-learn framework: Support Vector Machine (SVM) with a linear and rbf kernel, Random Forest (RF), Decision Tree (DT), and Logistic Regression (LR) with L1, L2, and ElasticNet regularization \cite{sklearn_api}. Due to the small size of our dataset, all experiments were conducted with 5-fold stratified cross-validation, and all models were implemented with default scikit-learn parameters, in order to avoid obtaining overly optimistic model performances \cite{Raschka2018ModelEM}. The cross-validation was conducted with a stratified approach to account for imbalanced amounts of data available across phonemes. All features in the training and testing set of each cross-validation fold were standardized per the distribution of the training set. Our primary metric was the phoneme classification accuracy, averaged across folds. A \textit{chance performance} baseline of 3\% (1$\slash$33) was used to compare our models to a random multi-class classifier. 


\section{Results and Discussion}
\subsection{Comparison of Speech Representations}
The average classification accuracy computed across all Ladin phonemes for each of 8 classifiers trained on 9 different speech representations is reported in \textbf{Table 1}. The highest classifier accuracy obtained was 86\%, achieved by a Dense NN trained on the segment-level MFCC speech representation. For each classifier, either the segment-level MFCC or WaveNet speech representations achieved the highest classification accuracy. 

For MFCC, the segment-level feature extraction approach always yielded a higher classification accuracy than the frame-level approach. We noted that the efficacy of the frame-level versus segment-level approach depended on the choice of underlying speech processing algorithm used; for LPC, the frame-level approach always yielded a higher classification accuracy than the segment-level approach. The poor performance of the segment-level LPC representation is likely due to the small size of the segments; we observed greater variance within frames for the segment-level LPC coefficients compared to the segment-level MFCC coefficients. 

Representations learned from the traditional autoencoders trained on the unlabeled Ladin speech files were outperformed by the segment-level MFCC and WaveNet speech representations. We noted that Big AE always outperformed Big LSTM AE across all classifiers. This is likely due to the limited amount of training data and the larger number of trainable parameters for LSTM networks. Small AE and Small LSTM AE exhibited similar performance across all classifiers. The small autoencoders always outperformed their larger counterparts, indicating that they learned more useful, condensed representations of phonemes in our dataset's context.  
  
\subsection{Comparison of Phoneme Classifiers}
We noted that the Dense NN achieved the highest accuracy across all speech representations.  These results suggest that deep neural network models for phoneme classification can outperform classical machine learning algorithms, even in low-resource language settings. In contrast, we noted that the linear SVM achieved the lowest accuracy across all speech representations. However, using an RBF kernel substantially increased SVM classification accuracy across all speech representations.    

\subsection{Phoneme Classification Performances}
\label{ssec:phoneme group comparison}
We evaluated the ability of Dense NN trained on MFCC segment representations to classify individual Ladin phonemes and phoneme subgroups, documented in \textbf{Table 2}. All phonemes and phoneme subgroups had  classification accuracies substantially higher than our chance level baseline (3\%), demonstrating the potential for leveraging discriminative MFCC segment representations to classify Ladin phonemes.

The average vowel classification accuracy (88\%) was higher than that of consonants (83\%), and average unrounded vowel classification accuracy (90\%) was higher than that of rounded vowels (80\%). The performance difference between unrounded and rounded vowels may have been influenced by the impact of rounding on speech formants: rounding lengthens the vocal tract, lowering all formants on the spectral envelope \cite{Zsiga2013TheSO}. Since MFCC represents spectral envelopes of sounds, features extracted from rounded vowels may have been less discriminative. Within the consonants, fricatives and nasals had the highest average classification accuracy (87\%), followed by plosives and trills (82\%), affricates (80\%) and approximants (77\%). Affricates are more temporally complex sounds, as they combine a fricative and plosive, which may contribute to their lower performance. Approximants form during slight constrictions of the vocal tract that are not as prominent or turbulent as constrictions used to produce fricatives; the complex and subtle nature of approximants may contribute to their lower performance \cite{Zsiga2013TheSO}. \textit{Our results for Ladin phoneme subgroup classification demonstrate the potential for using MFCC segment-level approaches to compute discriminative representations of Ladin across subgroups of speech sounds}. These findings motivate further development of techniques to improve the classification accuracy of more complex Ladin phoneme groups.

An automated phoneme classification system deployed in-the-wild to help linguists rapidly transcribe Ladin must perform effectively on individual phonemes, in addition to phoneme subgroups. Six individual phonemes exhibited classification accuracies above 90\%:  the unrounded vowels \textit{i}, \textit{e}, and \textit{a}, the fricatives \textit{\textipa{S}} and \textit{s}, and the nasal \textit{m}. The lowest-performing individual phonemes were the rounded vowel \textit{œ}, the fricative \textit{z}, the unrounded vowel \textit{\textipa{E}},  the plosive \textit{b}, and the nasals \textit{\ipa{N}} and \textit{\ipa{M}}; the low performances of these individual phonemes, compared to the other phonemes in their subgroups, may have resulted from the imbalanced distribution of phonemes in the dataset.  \textit{Our results for individual Ladin phoneme classification demonstrate the potential for using MFCC segment-level approaches to learn discriminative representations of individual Ladin speech sounds}. 

\begin{table}[!t]
\centering
\caption{Average classification accuracy for each Ladin phoneme subgroup, the 6 phonemes with the highest performance, and the 6 phonemes with the lowest performance, obtained by Dense NN trained on MFCC Segment representations.}
\begin{tabular}{l|c}
\textbf{Phonemes} & \textbf{Classification Accuracy} \\ \hline
\multicolumn{2}{c}{\textbf{Phoneme Subgroups}} \\ \hline
All Vowels & 0.88 \\
Rounded Vowels & 0.80 \\
Unrounded Vowels & 0.90 \\
\hline
All Consonants & 0.83 \\
Affricates & 0.80 \\
Approximants & 0.77 \\
Fricatives & 0.87 \\
Nasals & 0.87 \\
Plosives & 0.82 \\
Trills & 0.82 \\ \hline
\multicolumn{2}{c}{\textbf{Highest Phoneme Performances}} \\ \hline
i & 0.95 \\
e & 0.93 \\
\ipa{S} & 0.93 \\
a & 0.90\\
s & 0.90 \\ 
m & 0.90 \\ \hline
\multicolumn{2}{c}{\textbf{Lowest Phoneme Performances}} \\ \hline
œ & 0.50 \\
z & 0.62 \\
\ipa{E} & 0.63 \\
b & 0.67 \\
\ipa{N} & 0.67 \\
\ipa{M} & 0.67 \\
\end{tabular}
\end{table}

\section{Conclusions}
Our paper presents the first analysis of speech representations and machine learning models for classifying 32 phonemes of Ladin. We demonstrate the potential for learning discriminative representations of Ladin phonemes through traditional speech processing algorithms (MFCC in particular) and autoencoder-based approaches. We provide a proof-of-concept to recruit linguists to further transcribe recordings in our novel Ladin dataset. 

Future research includes experimenting with data augmentation \cite{ragni2014data} to increase the quantity of Ladin training data and developing multilingual models that learn cross-lingual phoneme representations to improve Ladin phoneme classifiers. Ladin shares phonetic and phonological attributes with non-endangered languages (e.g., Italian) \cite{yang_walker_vietti_chiocchetti_2021}, indicating that cross-lingual modeling may be a promising research direction for Ladin preservation. Future work also includes integrating our machine learning models into a real-world system to help linguists rapidly transcribe Ladin during field research (Figure 1). Our work contributes insights for the future development and deployment of phoneme classification approaches that can help linguists preserve Ladin and other low-resource endangered languages.

\section{Acknowledgements}
We thank the native speakers of Ladin for their generous participation in this research and contribution of recordings. We thank the Ladin Cultural Institute in Italy and the University of Southern California for supporting this research.

\section{References}
\renewcommand*{\bibfont}{\small}

\printbibliography[heading=header]


\end{document}